\documentclass[aps,showpacs]{revtex4}
\usepackage[T1]{fontenc}
\usepackage{wasysym}
\usepackage[dvips]{graphicx}

\newcommand{\ana}{{\it Astr. Astrophys.}}
\newcommand{\apjl}{{\it Astrophys. Jour. Lett.}}
\newcommand{\apjs}{{\it Astrophys. Jour. Suppl. Ser.}}
\newcommand{\mnras}{{\it Month. Not. Roy. Astron. Soc.}}
\newcommand{\aspc}{{\it Astron. Soc. Pac. Conf. Ser.}}
\newcommand{\cpc}{{\it Comp. Phys. Comm.}}
\newcommand{\zfa}{{\it Zeitschr. f. Astroph.}}

\newcommand{\msol}{M_{\astrosun}}
\newcommand{\lsol}{L_{\astrosun}}
\newcommand{\myr}{\mathrm{Myr}}
\newcommand{\mach}{\mathcal{M}}
\newcommand{\machmax}{\mathcal{M}_\mathrm{max}}
\newcommand{\machmean}{\mathcal{M}_\mathrm{mean}}
\newcommand{\ekin}{E_\mathrm{kin}}
\newcommand{\eint}{E_\mathrm{int}}
\newcommand{\msup}{M_\mathrm{sup}}
\newcommand{\mtot}{M_\mathrm{tot}}
\newcommand{\vturb}{v_\mathrm{turb}}
\newcommand{\cs}{c_\mathrm{s}}

\begin{document}

\title{Ionization front-driven turbulence in the clumpy interstellar medium}

\author{Thomas Peters\footnote{Fellow of the International Max Planck
Research School for Astronomy and Cosmic Physics at the University of
Heidelberg and the Heidelberg Graduate School of Fundamental Physics
}\footnote{Corresponding author: thomas.peters@ita.uni-heidelberg.de},
Robi Banerjee and Ralf S. Klessen}

\affiliation{Institut f\"{u}r Theoretische Astrophysik, Universit\"{a}t
Heidelberg, Albert-Ueberle-Str. 2, 69120 Heidelberg, Germany}

\begin{abstract}
We present 3D radiation-gasdynamical simulations of an ionization front
running into a dense clump. In our setup, a B0 star irradiates an overdensity
which is at a distance of $10\,\mathrm{pc}$ and modelled as a supercritical
$100\,\msol$ Bonnor-Ebert sphere. The radiation from the star
heats up the gas and creates a shock front that expands into the
interstellar medium. The shock compresses the clump material
while the ionizing radiation heats it up. The outcome of this
``cloud-crushing'' process is a fully turbulent gas in the wake of
the clump. In the end, the clump entirely dissolves. We propose that this
mechanism is very efficient in creating short-living supersonic turbulence
in the vicinity of massive stars.

\noindent
\end{abstract}

\maketitle

\section{Introduction}

Massive stars strongly influence the environment in which they have
formed by stellar winds, ionizing radiation and supernova explosions.
An ionization front which expands into the ambient interstellar medium (ISM)
and hits a dense clump may compress it so heavily that gravitational
collapse might be triggered. On the other hand, ionization heats
up the material and can photoevaporate the clump. The remaining material of this
competition may form low-mass stars~\cite{hest05} or brown dwarfs~\cite{whit04}.

The interaction of a shock with a dense clump is called ``cloud-crushing''.
The cloud-crushing scenario has been studied numerically both for
supernova shocks and ionization fronts. The first extensive studies
of the fate of the shocked cloud already showed that strong vortex rings can be
produced~\cite{klein94}. The mixing properties of the cloud depend
sensitively on the initial density distribution~\cite{naka06}.
Furthermore, simulations of dense clumps exposed to an ionizing flux
but without strong shocks show the generation of kinetic energy~\cite{kessel03}
and fragmentation of the clump~\cite{esquivel07}.
Radiation-gasdynamical simulations have also been used to match
observations in H II regions, especially the Eagle Nebula~\cite{williams01,
miao06}.

Although ionizing radiation injects a significant amount of energy
into the ISM, it does not seem to be an important driving mechanism
of interstellar turbulence on a global scale~\cite{maclow04}. However,
the cloud-crushing process generates a considerable amount of turbulence
locally in the wake of the cloud. We find that the motion of the cloud
material is mostly supersonic while the ambient gas behind the front
moves only subsonically. The continuous heating limits the lifetime of the
dense material, but the supersonic motions are maintained until the
cloud disperses. This is contrary to the situation in jet-clump
interactions, where the situation is less clear with some studies
showing mostly subsonic motions~\cite{baner07} while others claim
supersonic velocity fields~\cite{li07}.


\section{Numerical methods}

We perform 3D radiation-gasdynamical adaptive mesh simulations with the FLASH
code~\cite{fryxell00}, which is based on the PARAMESH library~\cite{macneice00}.
In addition to the refinement with respect to shocks we also make sure
that we resolve the Jeans length with a sufficient number of cells to avoid
artificial fragmentation~\cite{truelove97,baner04}. The Jeans length is the
critical scale for self-gravitating objects, and a dynamical increase of
resolution would indicate a runaway collapse. However, the run discussed
in this paper show collapse only temporarily.

Additionally to the standard hydrodynamics and self-gravity, we include the
ionization feedback from a massive star. The radiation feedback is included
via a raytracing approach~\cite{rijk06}. There are no radiation pressure terms
in the Euler equations; in this module coupling between hydrodynamics and
radiation takes place only through thermodynamics. We solve a rate equation
for the ionization fraction including collisional ionization and photoionization
as well as radiative recombination. The energy equation contains a
photoionization heating rate; the equation of state is isothermal.

\section{Setup}

The computational domain has dimensions
$(6.0 \times 2.4 \times 2.4) \cdot 10^{19}\,\mathrm{cm}$ with an effective
resolution of $512 \times 256 \times 256$ cells. We place a self-gravitating
supercritical gas sphere with Bonnor-Ebert density profile~\cite{bonnor56, ebert55}, with mass $M = 100\,\msol$ in
the centre of the domain. It slowly rotates with an angular velocity of
$\Omega = 7.83 \cdot 10^{-15}\,\mathrm{rad} / \mathrm{s}$ around the $z$-axis
corresponding to a ratio of rotational to gravitational energy of $\beta = 0.02$.
Additionally to the rotational velocity, a turbulent velocity field
with a magnitude of at most 50\% of the sound speed is added.
The temperature of the Bonnor-Ebert sphere is $T = 20\,\mathrm{K}$, while the
ambient gas has $T = 90\,\mathrm{K}$.

The ionizing source is located at the left hand side of the computational
domain at $(0.0, 1.2, 1.2) \cdot 10^{19}\,\mathrm{cm}$.
It has a temperature of $T_\mathrm{s} = 29,200\,\mathrm{K}$ and a luminosity
of $L_\mathrm{s} = 24,000\,\lsol$, representing a B0 star. The radiation
heats the interstellar gas to $T \approx 6 \cdot 10^4\,\mathrm{K}$.

\section{Results}

When the simulation starts, the gas next to the source becomes ionized
and heated. An ionization front accompanied by a shock expands into the
ISM. When the shock hits the clump, the clump material is swept away
and thereby compressed heavily. The shock leaves behind a fully turbulent
gas. The hot ionized gas is being mixed with the cold clump material while
the radiation heats it up continuously. After the shock front passes the
clump, the former clump disperses completely.

\begin{figure}
\begin{center}
\includegraphics{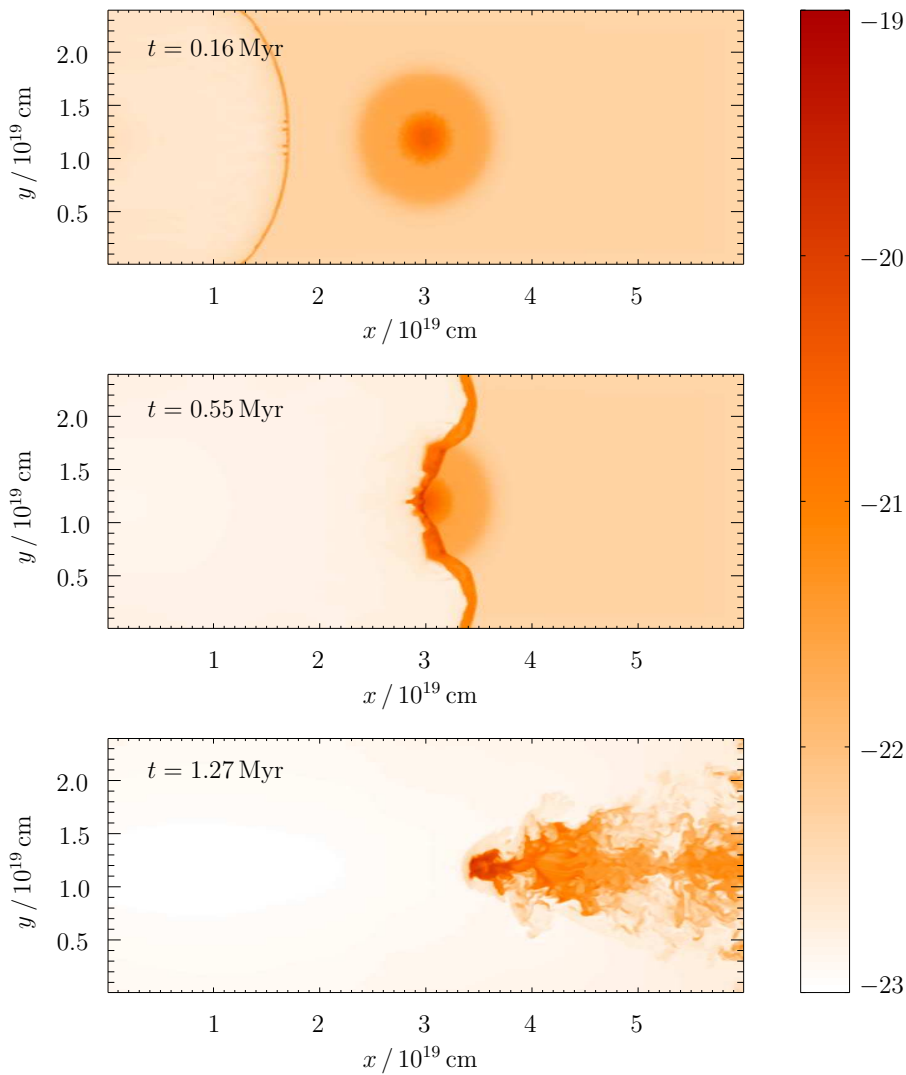}
\end{center}
\caption{The main stages of the simulation are shown as 2D cuts of the
logarithmic mass density $\log (\rho\,/\,\mathrm{g}\,\mathrm{cm}^{-3})$ in the
midplane. The different snapshots are taken at times $t = 0.16\,\myr$,
$t = 0.55\,\myr$ and $t = 1.27\,\myr$, respectively. The turbulent wake
behind the former clump at the last stage is clearly visible.}
\label{fig1}
\end{figure}

The time sequence of this process is as following. The shock touches the
outer edge of the clump at $t = 0.40\,\myr$. At $t = 0.53\,\myr$, it reaches
the clump centre at $x = 3\cdot10^{19}\,\mathrm{cm}$, and at $t = 0.76\,\myr$,
the front has totally enclosed the clump. Of course, the dense material delays
the shock, so that the wings of the front can propagate faster. They enter
the shadow zone and meet in the middle at $t = 0.77\,\myr$. The clash of these
wings is an important driving mechanism of the turbulence seen in these simulations.
It builds up an extended turbulent wake while the ionization front propagates further
into the ISM. At $t = 1.09\,\myr$, the shock reaches the boundary of the computational
domain at $x = 6\cdot10^{19}\,\mathrm{cm}$. The simulation stops at $t = 1.49\,\myr$
when the cloud has dissolved. Some of the stages of this time
evolution are depicted in figure~\ref{fig1}.

\section{Discussion}

\subsection{Energy balance}

\begin{figure}
\begin{center}
\includegraphics{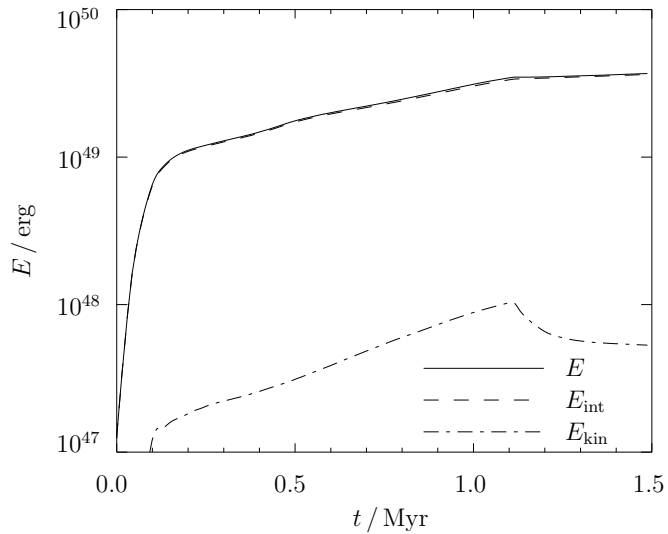}
\end{center}
\caption{The total energy $E$, internal energy $\eint$ and kinetic
energy $\ekin$ are shown as function of time $t$. The major contribution to
$E$ comes evidently from $\eint$, while $\ekin$ is between one and
two orders of magnitude smaller. The point in time where the shock, which
carries most of the kinetic energy, leaves the computational domain can
be distinguished easily.}
\label{fig2}
\end{figure}

We start our analysis with a brief look at the energy balance
in the flow (see figure~\ref{fig2}). The plot shows the total energy $E$,
internal energy $\eint$ and kinetic energy $\ekin$ in $\mathrm{erg}$ as a
function of time $t$. The B star provides energy to the system by ionising material.
This contribution is predominantly transferred into internal energy by the
photoionization heating, only a small fraction is converted into kinetic
energy. For example, at $t = 0.5\,\myr$, the ratio of internal
over kinetic energy is $\eint / \ekin \approx 57$.

Since the luminosity of the star is constant in time, the energy transferred
from the star to the gas in the computational domain grows linearly.
This explains qualitatively the form of $E(t)$ in figure~\ref{fig2}.
A quantitative analysis is difficult however, since geometric effects have to be taken
into account appropriately. One would have to account for the fact that
the star emits its radiation isotropically, while it is not at the centre
of a spherically symmetric computational domain, but on one face of
a Cartesian box.

\subsection{Mach numbers}

The simulation shows that the cloud-crushing flow is largely dominated
by supersonic motion. This is because the cloud material is cold, so that
the sound speed is much lower than in the hot gas behind the ionization
front, where a wind with $\mach \approx 0.2$--$0.4$ is observed. The cause
of the wind is that the photoionization heating is stronger close the source,
which leads to a pressure gradient and a corresponding flow. Since
the wind prevails in the largest part of the domain, namely the hot
postshock gas, the mean Mach number $\machmean$ is always below unity,
while the maximum Mach number $\machmax$, which is reached at crushing,
can be greater than $6$.

\begin{figure}
\begin{center}
\includegraphics{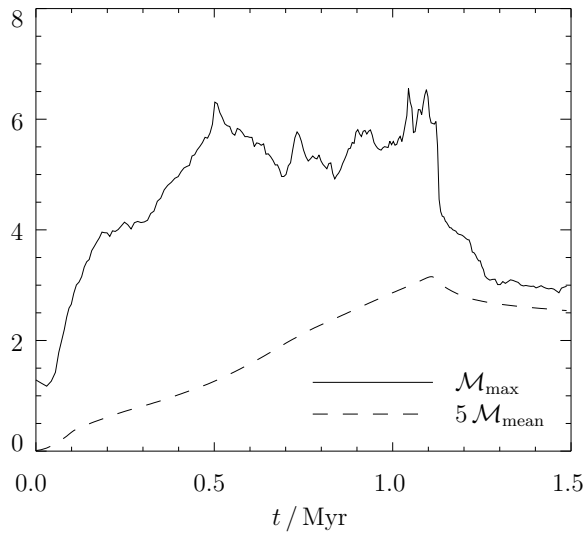}
\end{center}
\caption{The average and maximum Mach numbers in the flow provide information
on the dominance of shocks in the flow. To compare these two numbers, the
mean Mach number $\machmean$ is displayed amplified by a factor of $5$. The
peaks in $\machmax$ can be associated with events in the cloud-crushing
scenario.}
\label{fig3}
\end{figure}

Figure~\ref{fig3} depicts $\machmax$ and $\machmean$ as a function of time.
The maximum Mach number traces the shock ahead of the ionization front.
Within a time of $0.15\,\myr$, the shock accelerates up to a constant
velocity of $\mach = 4$. For a short time of another $0.15 \myr$ the
velocity seems to saturate, but the continuous photoionization heating
accelerates the shock again up to $\mach = 6.5$. At this point, when
$\machmax$ is maximal, the shock collides with the dense clump. As it hits
the high-density gas, the shock front moves more slowly. The gas decelerates,
so that $\machmax$ decreases again. However, the wings of
the shock that were not affected by the clump can enter the shadow
zone, which leads to a peak in $\machmax$ after $0.7\,\myr$. Then these
wings collide, which stops the motion in $y$-direction, so that the Mach
number decreases further. But the collision of the wings also leads to an
acceleration in positive $x$-direction, which can be seen in a series of peaks
from $0.8\,\myr$ to $1.1\,\myr$. After $1.1\,\myr$, the shock leaves
the computational domain, resulting in a sharp drop in $\machmax$.
This demonstrates that the highest Mach numbers are only reached in
the shock front itself, not in the shock-generated turbulence behind
the front. The motion in the turbulent wake is mostly supersonic with
$\mach$ below $3$. The mean Mach number grows continuously
until the shock leaves the domain, whereafter $\machmean$ declines slowly.
The heating of the gas does not change $\machmean$.

\begin{figure}
\begin{center}
\includegraphics{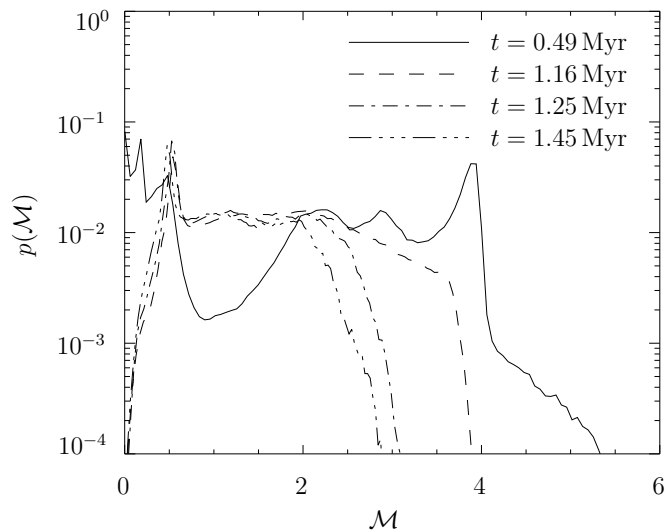}
\end{center}
\caption{The mass-weighted PDFs of the Mach number $\mach$ between times
$t = 0.49\,\myr$ and $t = 1.45\,\myr$. The shock front shows up as a peak
in the regime of high Mach numbers for the first point in time. At
$t = 1.16\,\myr$, the shock has left the computational domain, and the
turbulence with its supersonic Mach numbers decays.}
\label{fig4}
\end{figure}

The different stages of the simulation can also be recognized in the
probability density functions (PDFs) of the Mach number $\mach$. Figure~\ref{fig4}
shows mass-weighted PDFs at the moment of cloud-crushing ($t = 0.49\,\myr$)
and after the shock has left the domain ($t = 1.16\,\myr$ to $t = 1.45\,\myr$).
At the crushing time, the high $\mach$ above $2$ all belong to the shock.
The shock then excites supersonic turbulence in the wake, but away
from the shock $\mach$ above $3$ is very rare. While most of the
dense gas is supersonic, most of the domain is dominated by low
Mach number flows, both at crushing time and afterwards.

\begin{figure}
\begin{center}
\includegraphics{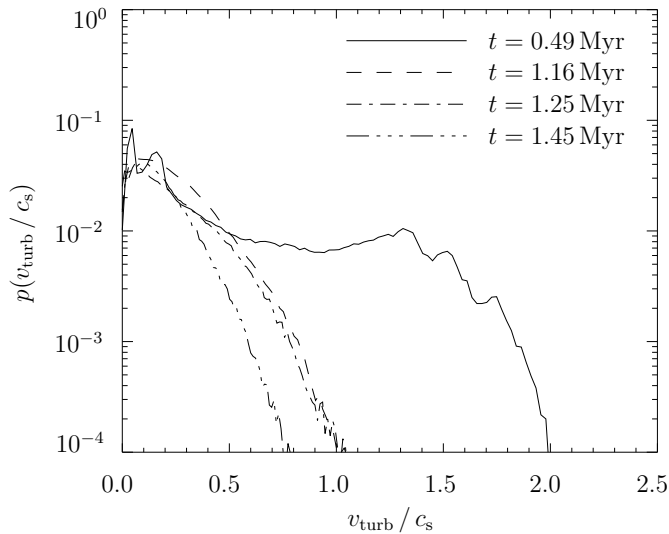}
\end{center}
\caption{The mass-weighted PDFs of the ``turbulent Mach number''
$\vturb / \cs$ with $\vturb = \sqrt{v_y^2 + v_z^2}$ for the same points in
time as in figure~\ref{fig4}. At cloud-crushing, the turbulent field $\vturb$
is slightly supersonic, while the flow in the wake is only subsonic. This
shows that the bulk motion contributes significantly to the high total
Mach numbers.}
\label{fig5}
\end{figure}

The PDFs of the total Mach number $\mach$ should be compared with
the Mach number given only by the turbulent velocity fluctuations,
$\vturb = \sqrt{v_y^2 + v_z^2}$, which is $\vturb /\cs$, where
$\cs$ is the local speed of sound. These plots
are shown in figure~\ref{fig5}. Since the bulk motion in
$x$-direction is no longer taken into account, the Mach numbers are
significantly lower. At the moment of cloud-crushing, the turbulent
Mach number is only slightly supersonic, while it is totally subsonic afterwards.
Hence, the bulk motion of the shock (and also the transport of momentum
by the wind) is important to reach the high Mach numbers observed
above.

\begin{figure}
\begin{center}
\includegraphics{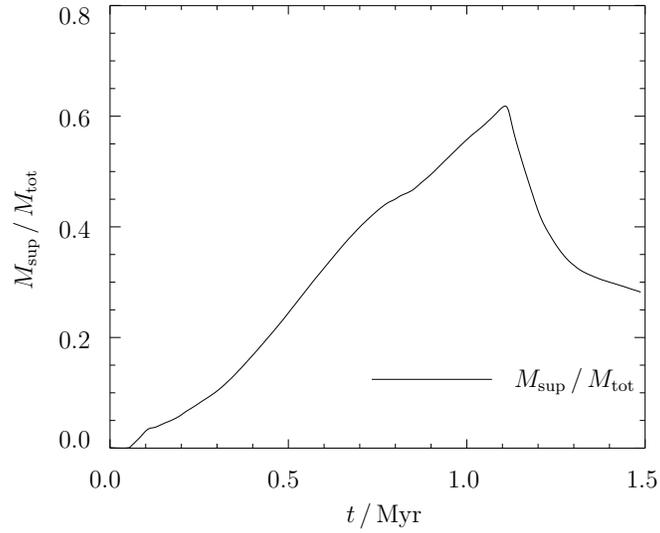}
\end{center}
\caption{The fraction of gas in supersonic motion $\msup / \mtot$
as a function of time $t$. Although most of the energy
input goes into internal energy, more than half of the gas moves
supersonically when the shock leaves the domain.}
\label{fig6}
\end{figure}

We are also interested in the fraction of mass which moves supersonically.
In figure~\ref{fig6} we plot the ratio of the mass of supersonic gas
$\msup$ and the total gas mass in the computational domain $\mtot$.
Despite of the complicated mixing processes,
$\msup / \mtot$ grows roughly linearly until the shock leaves the
domain. It is surprising that a significant part of the gas moves
supersonically, altough most of the energy input is converted into
internal energy (see figure~\ref{fig2}). When the shock
front leaves the domain, more than $60\%$ of the gas in our domain
is in supersonic motion.

\subsection{Clump structure after crushing}

In figure~\ref{fig7}, we enlarge a part of figure~\ref{fig1} belonging
to the snapshot at $t = 1.27\,\myr$ and show additionally to the mass
density also the temperature and the velocity components $v_x$ and $v_y$.
The remains of the dense core have a temperature around
$T \approx 300\,\mathrm{K}$, while the ambient ionized medium is at
$T \approx 3.5 \cdot 10^4\,\mathrm{K}$. The high temperatures in the
environment give rise to the ``rocket effect'', which accelerates the
gas in positive $x$-direction~\cite{oort55}. This is because the cold
gas at the surface of the clump facing the star becomes heated. Thus, it
expands into the postshock medium, carrying momentum with
it, and consequently the clump accelerates.

The cloud-crushing leads to vortical structures in the wake as can be
seen from the lower plots in figure~\ref{fig7}. The two largest structures
around $x \approx 4.2 \cdot 10^{19}\,\mathrm{cm}$ and $y \approx 0.7 \cdot10^{19}\,
\mathrm{cm}$ or $y \approx 1.7 \cdot 10^{19}\,\mathrm{cm}$, respectively, even
have slightly negative $v_x$ and so does a region around the tip of the
former core at $x = 3.3 \cdot 10^{19}\,\mathrm{cm}$ and $y = 1.2 \cdot
10^{19}\,\mathrm{cm}$. Averaged over the whole volume, however, the wind,
wich moves with $v_x \approx 1.5 \cdot 10^6\,\mathrm{cm}/\mathrm{s}$, causes a
bulk motion of the gas in positive $x$-direction. In order to measure the
turbulent components of the velocity field, it is better to focus on the
transversal directions. The peak amplitude of the velocity component in
$y$-direction, for example, is about half of the maximum velocity in
$x$-direction. The large vortical structures
discussed above have total velocities around $10^6\,\mathrm{cm}/\mathrm{s}$.
The upper vortex rotates clockwise, while the lower vortex rotates
counter-clockwise.

\subsection{Line-of-sight velocity profiles}

\begin{figure}
\begin{center}
\includegraphics{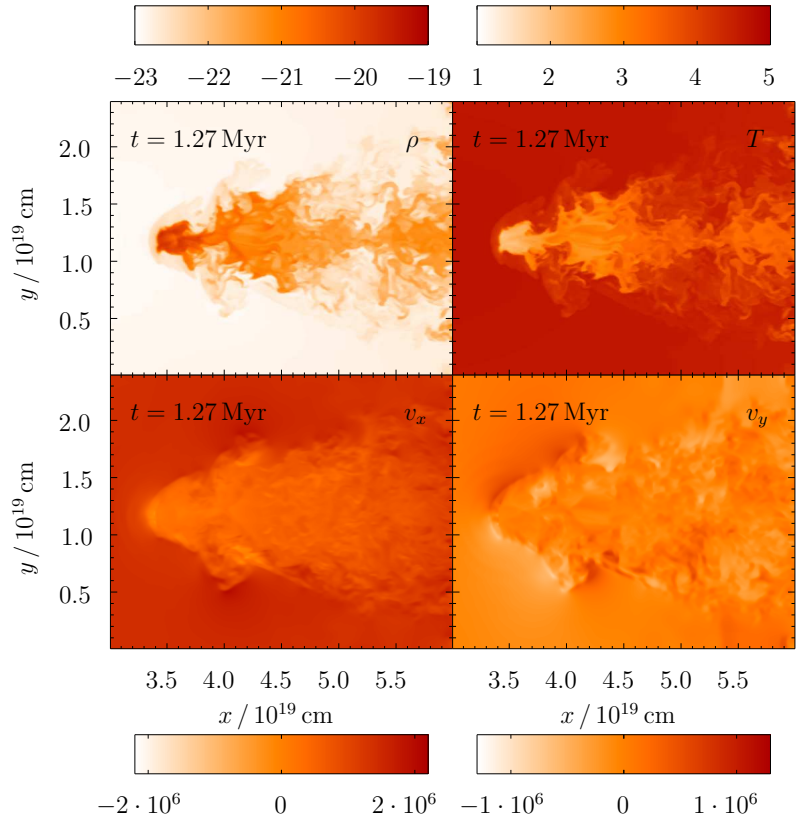}
\end{center}
\caption{The midplane cuts at $t = 1.27\,\myr$ of the density
$\log (\rho\,/\,\mathrm{g}\,\mathrm{cm}^{-3})$, the temperature
$\log (T\,/\,\mathrm{K})$, the velocity in $x$-direction
$v_x\,/\,\mathrm{cm}\,\mathrm{s}^{-1}$ and the velocity in $y$-direction
$v_y\,/\,\mathrm{cm}\,\mathrm{s}^{-1}$ are compared with each other.
The remains of the dense clump are still cold, while the material in
the wake is already heating up. While $v_y$ is a measure for the
turbulent velocity fluctuations, $v_x$  shows the bulk motion of the
gas.}
\label{fig7}
\end{figure}

\begin{figure}
\begin{center}
\includegraphics{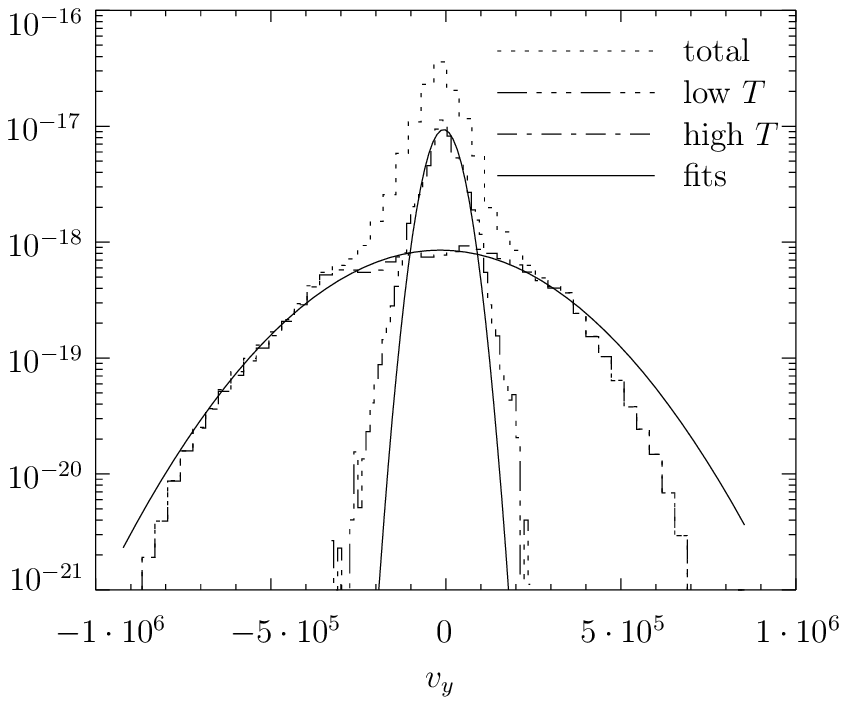}
\end{center}
\caption{Line-of-sight mass-weighted histograms for $v_y$ around the dense
core after cloud-crushing at time $t = 1.27\,\mathrm{Myr}$ (in
arbitrary units). The figure
shows both the total data and the low temperature $T \leq 10^3\,\mathrm{K}$
and high temperature $T \geq 10^4\,\mathrm{K}$ cuts. Clearly, the high
velocity contributions stem from the high $T$ gas, whereas the peak around
zero relates mainly to the low $T$ gas and a warm envelope with $T$ between
the two cuts. Also shown are Gaussian fits
to the low and high $T$ cuts that give insight into the strength of
the turbulent velocity fluctuations and can be compared with observed
spectral line widths.}
\label{fig8}
\end{figure}

A connection to observations of molecular clouds can be made by looking
at velocity profiles along a certain line-of-sight~\cite{ossen02}. Examplary, we show
in figure~\ref{fig8} a profile for $v_y$ of the dense core after crushing at time
$t = 1.27\,\mathrm{Myr}$ (compare also figure~\ref{fig7}). The width of
the beam is $3.0 \cdot 10^{19}\,\mathrm{cm} \leq x \leq
4.0 \cdot 10^{19}\,\mathrm{cm}$ and $0.7 \cdot 10^{19}\,\mathrm{cm} \leq z
\leq 1.7 \cdot 10^{19}\,\mathrm{cm}$, while $y$ ranges over the
whole box width. Note that the profile is a mass-weighted histogram, not a normalized PDF.
We do the same calculation again, but this time only considering gas
that has $T \leq 10^3\,\mathrm{K}$ (low $T$ case) or
$T \geq 10^4\,\mathrm{K}$ (high $T$ case). From the figure we see that
the high velocity tails are exclusively related to the high temperature gas.
The peak at low velocities comes mainly from both the low temperature
medium as well as a warm envelope with temperatures between the cuts
$10^3\,\mathrm{K} \leq T \leq 10^4\,\mathrm{K}$.
Additionally to the histograms, we have fitted Gaussians to the low and high
$T$ data. Their variance gives a measure for the turbulent velocity and can
be related to the width of spectral lines that are being oberserved.

\section{Conclusion}

We have seen that cloud-crushing by ionization fronts can lead to
short-living supersonic turbulence. Altough only a minute fraction
of the energy input is converted into kinetic energy, up to $60\%$
of the affected gas is supersonic. While it is mainly the cold gas that is
highly supersonic, it is the hot gas that moves the fastest. The bulk
motion of the shock is an important contribution to the supersonic flow,
since the transversal fluctuations are at best slightly supersonic.

\section*{Acknowledgements}

RB is funded by the Emmy-Noether grant (DFG) BA 3607/1.
The FLASH code was in part developed by the DOE-supported Alliances Center
for Astrophysical Thermonuclear Flashes (ASCI) at the University of Chicago.
TP wishes to thank the organizers of the International Conference ``Turbulent
Mixing and Beyond'' for their kind hospitality at the Abdus Salam International
Centre for Theoretical Physics and their financial support.

\end{document}